\documentclass[12pt
]{revtex4}

\usepackage{graphicx}

\begin{document}

\title{Bose-Einstein Condensation of Excitons in Cu$_2$O: Progress Over Thirty Years}

\author{David Snoke}

\affiliation{Department of Physics and Astronomy \\ University of Pittsburgh, Pittsburgh, PA 15260}

\author{G. M. Kavoulakis}

\affiliation{Technological Educational Institute of Crete \\
P. O. Box 1939, GR-71004, Heraklion, Greece}

\begin{abstract}
Experiments on Bose-Einstein condensation (BEC) of excitons in the semiconductor Cu$_2$O started over thirty years ago, as one of the first serious attempts at exciton BEC. Early claims were based on spectroscopic signatures and transport data which have since been reinterpreted, in large part because the Auger recombination process for excitons was not well understood. Understanding of the Auger process has also advanced, and recent experiments have made significant progress toward exciton BEC. We review the history of experiments on exciton BEC in Cu$_2$O, the Auger recombination process, and the prospects for observing exciton BEC in this system in the near future.
\end{abstract}


\maketitle

\vspace{1cm}

\section{Introduction}

{\bf BEC of excitons}. Bose-Einstein condensation (BEC) of excitons was predicted in the early 1960's, by Moskalenko \cite{mosk1}, Blatt and coworkers \cite{blatt}, and Casella \cite{casella}, in independent efforts. The concept is simple: just as two electrons with half-integer spin can be paired to make a Cooper pair which acts as an integer-spin boson, and just as an even number of fermionic electrons, protons, and neutrons are bound together to make up every bosonic atom, so too an electron and hole in a semiconductor, each with half-integer spin, can be bound together in the complex known as an exciton, which will also act as a boson. Therefore one can expect Bose-Einstein condensation of excitons under two conditions: 1) when the number of excitons is approximately conserved; in particular, their lifetime is much longer than the time it takes for them to thermalize to a well-defined temperature, and 2) when the density of the excitons is not too high, because competing phases occur at high density such as electron-hole plasma, in which the binding of the electrons and holes into excitons breaks down. 

Under these conditions, it is straightforward to describe the exciton gas by weakly interacting Bose gas theory. The general condition for quantum coherent effects to be important is that the thermal deBroglie wavelength $\lambda_{dB}$ of the particles be comparable to or larger than the average distance $r_s$ between the particles. The order of magnitude of the deBroglie wavelength is found by equating
\begin{equation}
\frac{\hbar^2k^2}{2m} = \frac{\hbar^2 (2\pi)^2}{2m\lambda_{dB}^2} \sim  k_BT,
\end{equation}
which implies
\begin{equation}
\lambda_{dB} \sim \frac{2\pi\hbar}{\sqrt{2mk_BT}}.
\end{equation}
In three dimensions, the average distance $r_s$ scales as $r_s \sim n^{-1/3}$, where $n$ is the density of the particles. Setting $\lambda_{dB} \sim r_s$ gives 
\begin{equation}
n \sim \frac{2^{3/2}}{(2\pi)^3}\frac{(mk_BT)^{3/2}}{\hbar^3},
\label{nc}
\end{equation}
or 
\begin{equation}
T \sim \frac{(2\pi\hbar)^2}{2mk_B}n^{2/3}.
\label{Tc}
\end{equation}
In equilibrium, the standard calculation of statistical mechanics \cite{statmech} gives the critical density for Bose-Einstein condensation,
\begin{equation}
T_c = 0.17 \frac{(2\pi\hbar)^2}{2mk_B}n^{2/3}.
\end{equation}
Even away from perfect equilibrium, however, the relations (\ref{nc}) and (\ref{Tc}) will still apply as the conditions for Bose-Einstein statistical effects, which include a peaking of the distribution of the particles near the ground state, also known as a ``quasicondensate.''

Note that these relations give a critical temperature inversely proportional to the mass $m$. This means that light mass implies Bose-Einstein effects at higher temperature. Since excitons have mass of the order of two electron masses, this led to the proposal that exciton condensates could be the first room temperature condensate. The excitons in many semiconductors are stable at room temperature, e.g. Cu$_2$O, CuCl, CdSe, ZnO, and GaN.  To have a Bose-Einstein condensate, one must ensure that the exciton lifetime is also long enough to allow thermalization, as discussed above. Several of these semiconductors seemed to satisfy this requirement, and experiments on them began in the 1970's, as discussed below.
\begin{figure}
\includegraphics[width=0.7\textwidth]{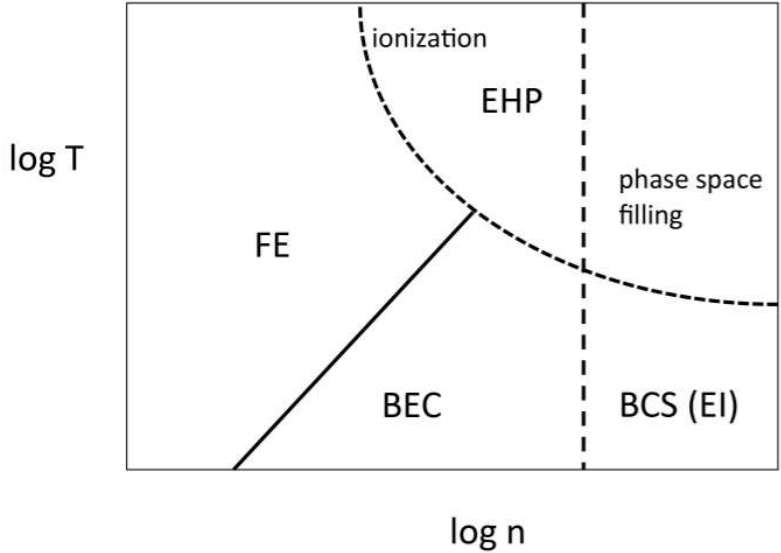}
\caption{Generic phase diagram of excitons in a three-dimensional bulk semiconductor. FE = free exciton gas. BEC = exciton BEC. BCS (EI) = excitonic insulator state, which is a BCS state; there is a BEC-BCS crossover at low temperature as density increases. EHP = electron-hole plasma, i.e. ionized excitons. There are two regimes for this, a non-degenerate and a degenerate plasma. The general condition for degeneracy is $r_s \sim a$, that is, the average distance between the particles be comparable to the exciton Bohr radius.}
\label{phasediagram}
\end{figure}

The relations  (\ref{nc}) and (\ref{Tc}) also imply that the critical temperature for Bose-Einstein effects depends on the density. In liquid helium and in BCS superconductors, the density is essentially a fixed quantity. By contrast, it is quite easy to vary the density of excitons to very low density by simply changing the intensity of a light source which generates electrons in the conduction band and holes in the valence band of the semiconductor. Therefore one has a density-dependent critical temperature, $T_c \sim n^{2/3}$, as plotted in Fig.~\ref{phasediagram}.  From the ideal Bose gas theory, one would say that to have a condensate at room temperature, one simply needs to increase the density of the excitons until their critical temperature for condensation is above room temperature. However, there are limits to the density of excitons which can be obtained. These limits come from 1) competing phases of the electrons and holes, and 2) density-dependent recombination mechanisms. 

{\bf Competing phases}. There are three possible competing phases for electrons and holes besides the exciton Bose-Einstein condensate (EBEC). One of these is the ``excitonic insulator'' (EI) state which is a BCS state of the electrons and holes instead of two spin states of the electrons. This state arises under the general condition $r_s < a$, where $a$ is the exciton Bohr radius, i.e, the pair correlation length. This is equivalent to the condition $na^3 > 1$. This condition corresponds to the case that a Fermi level of the electrons (and also of the holes, assuming they have comparable mass) exists in the system which is large compared to the intrinsic binding energy of the excitons. We can see this by noting first that the exciton binding energy ${\rm Ry}_{ex}$ and the excitonic Bohr radius $a$ are related by the formula (in MKS units) 
\begin{eqnarray}
{\rm Ry}_{ex} = \frac{e^2}{8\pi\epsilon a},
\end{eqnarray}
which is the same as for a hydrogen atom except that the electric force has been renormalized by the screening in the medium, $e^2 \rightarrow e^2/\epsilon$, where $\epsilon$ is the dielectric constant of the medium. The hydrogenic formula for $a$ is 
\begin{equation}
a = \frac{4\pi\epsilon\hbar^2}{e^2m}.
\label{ebohr}
\end{equation}
On the other hand, the Fermi energy of an electron gas in the low-temperature limit is given by \cite{statmech}
\begin{equation}
E_F = \left(\frac{3\pi^2\hbar^3}{\sqrt{2}m^{3/2}} n\right)^{2/3},
\end{equation}
where $n$ is the density. Setting this larger than the binding energy gives
\begin{eqnarray}
\frac{3\pi^2\hbar^3}{\sqrt{2}m^{3/2}} n &>& \left(\frac{e^2}{8\pi\epsilon a}\right)^{3/2} \nonumber \\
\sim \left(  \frac{4\pi\epsilon \hbar^2}{e^2m} \right)^{3/2} n &>& \frac{1}{a^{3/2}},
\end{eqnarray}
which is equivalent to $na^3 > 1$, using (\ref{ebohr}) for the excitonic Bohr radius.

In the EI limit the gas still is coherent, like a BCS superconductor, below a critical temperature. There should be a crossover from exciton BEC to the excitonic insulator state at low temperature, as density is increased. Indeed, some of the earliest work on the theory of BEC-BCS crossover was done in the context of thinking about exciton systems, notably foundational theory by Keldysh and Kozlov \cite{keld,keld2}, Hanamura and Haug \cite{HH1,HH2,HH3}, and Comte and Nozi\'eres \cite{comnoz1,comnoz2}; for a general review of the early theory of excitonic condensates, see Ref.~\cite{moskbook}). These works showed that excitonic condensation in the low-density BEC limit was a sound concept, despite the fermionic nature of the underlying electrons and holes, and there is a smooth transition to the BCS-like EI phase at high density. The critical temperature for this high-density EI phase will decrease with increasing density, however, so that raising the density is no longer advantageous. In practical terms, this means that the density of the particles should be kept below $1/a^3$. Excitonic condensation can be expected at room temperature only if $a$ is small enough that this limit can be satisfied even for densities given by (\ref{nc}) at room temperature. The condition that the critical density at room temperature be less than $1/a^3$ turns out not to be an issue for most of the semiconductors listed above with excitons which exist at room temperature, since large binding energy of the excitons also corresponds to small Bohr radius.

Another competing phase is ionization to an incoherent electron-hole plasma (EHP). This is analogous to the classical ionization of an atomic gas at high temperature and high density due to three-body collisions; it is sometimes called the excitonic ``Mott'' transition because it is a conductor-insulator transition, although the mechanism is quite different from the Mott transition in cold-atom lattices. The theory of this transition is actually quite complicated (for a review, see Ref.~\cite{exion}; see also recent work by Manzke and coworkers \cite{manzkenew1,manzkenew2}). Self-consistent theory of this transition involves the dynamic screening of the electron-hole interaction, which in turn depends on the number of ionized electrons and holes. Figure~\ref{phasediagram} illustrates the basic shape of the boundary for this phase. The position of the phase boundary depends on the binding energy of the excitons; in general we can say that higher binding energy pushes this curve to higher temperature. Therefore  having deeply bound excitons also helps avoid this competing phase.

Another competing phase is the electron-hole liquid (EHL), sometimes also called ``condensation.'' This state received a significant amount of attention in the 1970's and 1980's \cite{EHLexpt2,EHLtheory-keld} following its observation in two bulk semiconductors, Ge and Si. This state is  analogous to liquid mercury: the electrons and holes generated by optical excitation are not bound into pairs as excitons or biexcitons but instead form two interpenetrating Fermi gases with the properties of a classical liquid, with a surface tension. The EHL state is a conductor, while the EBEC state is an insulator, because excitons are charge neutral. If the EHL state does exist, it will prevent EBEC at any temperature and density, since its phase boundary also scales as $T \sim n^{2/3}$ in three dimensions. 

The EHL state only occurs in indirect-gap semiconductors like Si and Ge with multiple degenerate valleys in the conduction band. This degeneracy allows the density of the excited electrons to be high while still keeping the average kinetic energy of the carriers low. This means that the mutual attraction of the electrons and holes overcomes the kinetic energy cost of the Fermi level of free carriers, leading to a net free energy savings to enter the EHL state at high excited carrier density \cite{EHLtheory-keld}. In a nice experiment, Timofeev and coworkers \cite{timo-Ge} used a stress geometry to lift this degeneracy in a germanium crystal, and magnetic field to prevent biexciton formation, and found that the EHL state did not occur, leaving only free excitons. At high density they saw evidence for Bose-Einstein statistics of the excitons (see Ref.~\cite{moskbook}, section 1.4.2), although not for EBEC. 

As illustrated in Fig.~\ref{phasediagram}, it is expected that at low enough temperature and density, excitons can undergo BEC, and if the excitons have large binding energy and small Bohr radius this could even occur at room temperature. Therefore 
experimental research on Bose-Einstein condensation of excitons began in the 1970's with a focus on deeply bound excitons in bulk semiconductors.  As we will see, however, the second competing effect mentioned above, namely density-dependent recombination mechanisms, have turned out to be a major issue.

{\bf Deeply bound excitons and Cu$_2$O}. The earliest experiments on excitonic condensation in bulk semiconductors were done with the bulk semiconductors CdSe and CuCl \cite{japan1,japan2,japan3,japan4}. The most complete study on CuCl was done in the early 1980's  \cite{peygh1,peygh2}. In CuCl, there is a tightly bound biexciton state (i.e., an excitonic molecule like H$_2$), which, of course, is also an integer-spin boson which can in principle undergo BEC. The biexciton binding energy in CuCl is 26 meV, which is comparable to $k_BT$ at room temperature. The experiments with CuCl ran into interpretive difficulties, however, because CuCl has a strong polariton effect, in which photons in the medium couple strongly to the exciton or biexciton states, leading to a very short radiative lifetime for low-momentum excitons. At high density, the biexcitons in CuCl showed a distinct peak in low-energy states, but were far from equilibrium.  The enhancement of the population in low-energy states was probably related to the boson nature of the particles through the process of stimulated scattering, which drives the onset of BEC in all bosonic gases \cite{yl,snoke-wolfe}. In recent years, several sophisticated models \cite{hartwell,tassone,malpuech} have been developed which describe the nonequilibrium behavior of  polaritonic gases and show clearly the role of the bosonic stimulated scattering. It would be quite interesting to apply these modern theoretical tools to the early CuCl experiments to estimate to what degree the occupation number of the biexcitons was enhanced by their bosonic nature. 

Around the same time, the semiconductor Cu$_2$O was identified as a good candidate for excitonic BEC. The excitons in Cu$_2$O have binding energy of 150 meV \cite{binding}; the biexciton state is either nonexistent or only very weakly bound, on the order of 1 meV or less \cite{cu2obiex,mosk-biex}, so that one can generally ignore the existence of biexcitons (we discuss this further below). 

The lowest conduction band in Cu$_2$O comes from the $4s$-orbitals of the oxygen atoms, while the highest valence band comes from the $3d$-orbitals of the copper atoms. The crystal field in the cubic lattice splits the five $d$-orbitals into a triplet and a doublet, with the triplet higher; when the electron spin is taken into account, the spin-orbit interaction splits this triplet into a higher twofold-degenerate state and a lower fourfold-degenerate state. The lowest-energy excitons are formed from holes in these two highest states of the valence band and electrons in the two $s$-like states of the lowest conduction band.  The electron-hole exchange splits these four states into a triplet and a singlet, which are called the orthoexciton and the paraexciton, respectively. These are split by 12 meV, with the paraexciton the lower state. In general, these two exciton states, which are the lowest energy exciton states, are the only important ones unless higher-lying exciton states are being excited directly. 

The symmetry group of unstressed Cu$_2$O is $O_h$, which is cubic with inversion symmetry. This high symmetry leads to the effect that the lowest exciton state in Cu$_2$O, the singlet paraexciton has zero oscillator strength for interaction with photons, due to the selection rules which arise from the symmetry of the crystal \cite{binding,moskbookA,bayer}. This implies that there is no polariton effect for the paraexcitons. The triplet orthoexciton state, by contrast, has a symmetry-allowed radiative recombination process. 
(The singlet-triplet terminology of Cu$_2$O sometimes confuses those who work with organic light emitters, in which the state with forbidden recombination is the triplet state and the state with allowed radiative recombination is the singlet state. The difference in the terminology is because in the case of organics, the transition is assumed to have an $s$-$p$ dipole-allowed matrix element, and the spin singlet state corresponds to no spin flip of the excited electron, and thus an allowed process, while the spin triplet state corresponds to a spin flip. In the case of Cu$_2$O, neither the orthoexciton nor the paraexciton is a pure spin state; the triplet orthoexciton state is actually a superposition of singlet and triplet pure electron spin states, while the singlet paraexciton is a superposition of only pure spin triplet states.)

In principle, the fact that the lowest exciton state has a forbidden transition means that these excitons have infinite lifetime, but a weak phonon-assisted recombination process is allowed \cite{birman}, and also 
recombination at impurities can occur, so that the lifetime of paraexcitons in Cu$_2$O is typically of the order of hundreds of nanoseconds to microseconds \cite{andre,snoke-shields,naka-njp}. This is very long compared to excitons in most semiconductors, and very long compared to the thermalization times of the excitons via exciton-phonon interaction, which is tens of picoseconds \cite{snokebc}. Because there has never been a large technological effort to produce pure Cu$_2$O,  most of the purest samples come from rock collectors who obtain them from copper mines. 

A very important step was accomplished in the 1980's in the trapping of excitons in Cu$_2$O. Inhomogeneous stress was used to create a harmonic potential minimum for the excitons inside the crystal \cite{trauer,trauer2} very similar in geometry to the magneto-optical traps used to hold cold atoms for BEC experiments. As discussed below, some early experiments on EBEC used surface excitation of the crystal, which had the advantage of achieving high carrier density due to the short absorption length of the excitation light, but generated a highly nonequilibrium situation with an expanding gas that was difficult to analyze. The stress used in the exciton traps has the side effect that it changes the symmetry of the crystal, so that the paraexciton state no longer has a forbidden radiative recombination transition \cite{parastress}, but the lifetime of the paraexcitons in a stressed crystal is still very long compared to the exciton-phonon interaction time. 

\begin{figure}
\includegraphics[width=0.5\textwidth]{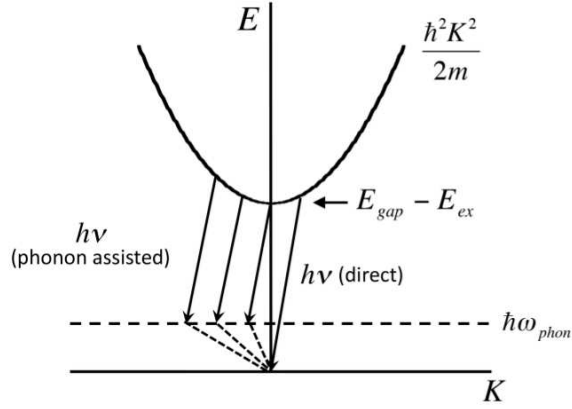}
\caption{Recombination processes of Wannier excitons in a direct band-gap
semiconductor. $K$ is the wave vector associated with the center-of-mass momentum of the exciton; $E_{gap}$ is the semiconductor band gap energy, and $E_{ex}$ is the exciton binding energy. $\hbar\omega_{phon}$ is the energy of an optical phonon, which is nearly independent of momentum.}
\label{fig.pha}
\end{figure}
{\bf Phonon-assisted luminescence spectroscopy}. The phonon-assisted recombination process in Cu$_2$O provides nice data to indicate the behavior of the excitons and whether they undergo BEC. Figure~\ref{fig.pha} shows the phonon-assisted recombination process of excitons in a direct-gap semiconductor like Cu$_2$O. Only excitons with low momentum can recombine via the direct recombination process, while an exciton at any momentum can recombine via the phonon-assisted process, with the phonon taking up any excess momentum. 
The phonon-assisted process can occur for both the paraexciton, which is a singlet state, and the orthoexciton state, which is a triplet that lies 12 meV above the paraexciton in unstressed crystals.  The energy of the emitted photon is equal to the total energy of the exciton minus the energy of the optical phonon, which is nearly constant. The energy spectrum of the phonon-assisted luminescence therefore gives the kinetic energy distribution of the excitons directly. If we take the matrix element as nearly independent of the exciton momentum, which is the case in Cu$_2$O, then the intensity of the light emitted at a given energy is directly proportional to the number of excitons with the corresponding kinetic energy $\hbar^2K^2/2m$.
\begin{figure}
\includegraphics[width=0.46\textwidth]{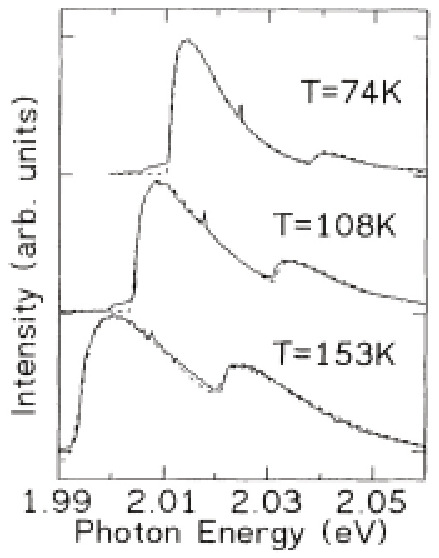}
\includegraphics[width=0.45\textwidth]{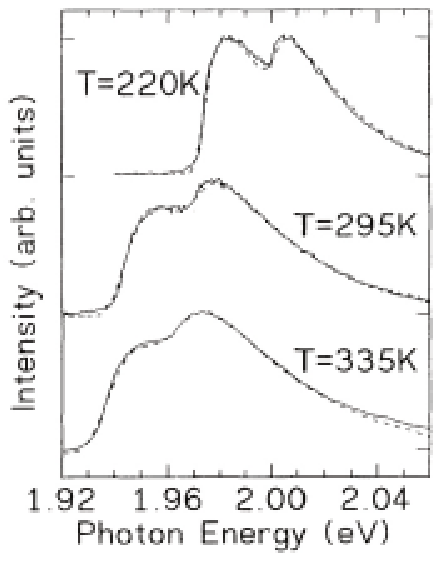}
\caption{Solid lines: luminescence from excitons in Cu$_2$O from low temperature to room temperature. Dashed line: fit to the theory for phonon-assisted luminescence discussed in the text, giving the thermal distribution of the excitons. A single parameter, namely the lattice temperature, is used to fit all the curves.  From Ref.~\protect\cite{snoke-shields}.}
\label{figRT}
\end{figure}

Figure~\ref{figRT} shows typical spectra of phonon-assisted luminescence from orthoexcitons in Cu$_2$O at various temperatures up to room temperature. At high temperature, an optical phonon can be both emitted or absorbed when an exciton recombines. Therefore there are two lines, one at the exciton energy minus the optical phonon energy, and one at the exciton energy plus the optical phonon energy.  The theory fits to each peak in the data in Fig.~\ref{figRT} are given simply by $I(E+E_0) \propto D(E)f(E)$, where $E$ is the exciton kinetic energy, $D(E)$ is the density of states of the excitons, proportional to $E^{1/2}$ in three dimensions, and $f(E)$ is the occupation number of the excitons, proportional to $e^{-E/k_BT}$ at low density.  The relative height of the two lines depends on the temperature of the lattice. At low temperature, the phonon-emission line dominates, since there are no phonons to absorb, while at high temperature, the two lines become comparable.

When many-body interactions are taken into account, the analysis of the phonon-assisted luminescence is slightly more complicated. Each momentum state emits with an energy profile given by the spectral function $A(\vec{k},\omega)$, which is approximately equal to $\delta(\omega-\omega_{\vec{k}})$ at low density and low temperature, but can be substantially broadened when there are strong particle-particle interactions \cite{snokebook,griffinshi,zimm}. This brings up a crucial difference between exciton BEC experiments and experiments on cold atoms.  The fact that excitons can recombine and turn into photons allows us to have direct access to the spectral function. (This property has been used widely in the exciton-polariton BEC experiments, e.g. to show spectral narrowing due to coherence \cite{polspec1,polspec2}.) In cold atom experiments, the spectral function is a theoretical construct that is never directly observed. 

Direct access to the spectral function means that we can see absolute shifts of the ground-state energy of the excitons directly as energy shifts of the emitted photons, and that we can see the effect of broadening of the spectral function as spectral broadening of the photon emission. To lowest order, the former can be identified as the real part of the exciton self-energy, and the latter as the imaginary part of the self-energy. (See Ref.~\cite{snokebook}, section 8.4). The exciton-phonon interaction increases as the temperature increases, leading to a red shift of the luminescence line in Fig.~\ref{figRT} due to renormalization of the real self-energy as well as broadening due to increasing imaginary self-energy. The imaginary self-energy is, to first order, just given by $\Gamma = \hbar/\tau$, where $\tau$ is the average scattering time of the excitons. This scattering time is an average of all scattering processes, including exciton-phonon scattering, which increases as temperature increases, exciton-impurity scattering, which also increases with increasing temperature, and exciton-exciton scattering.

{\bf Exciton-exciton scattering}.  This line broadening effect was used in the first attempt to estimate the absolute density of the excitons independently from the assumed statistics of the excitons in the Cu$_2$O experiments \cite{comments}, as discussed below. Given an estimate of the exciton-exciton scattering 
cross section, the scattering time $\tau$ for a given density can be calculated for the case when the lattice is cold, near 1 K, so that exciton-phonon interactions are negligible, and this can be compared to the observed broadening as a function of exciton density.

The cross section for exciton-exciton scattering is actually a problematic theoretical topic for exciton physics, as compared to cold atom physics, where atom-atom cross sections are calculated to three decimal places or better. The difficulty of calculating the exciton-exciton scattering cross section has nothing to do with details of the band structures of the crystals, which are well understood. The primary difficulty is the fact that the electron and hole have comparable mass, so that all the following exchange effects are important and cannot be neglected: electron-electron exchange, hole-hole exchange, and electron-hole exchange. In atoms, there is no electron-nucleus exchange, and nucleon-nucleon exchange can typically be ignored (although M. Combescot  has argued that under the conditions of BEC the nucleon-nucleon exchange may be important, since the wavelength, by definition, is comparable to the distance between the particles; see Ref.~\cite{comb} and references therein for the general approach of accounting for these exchange effects). 

The most accurate calculations for the exciton-exciton scattering cross section in Cu$_2$O have been done by Shumway and Ceperley \cite{shum}, and give a cross section of 5~\AA, which is comparable to the Bohr radius of the excitons. In that work, the spin of the electrons and holes was assumed to be aligned. The spin structure of the excitons can lead to significant differences in the scattering cross sections between different spin states \cite{mosk-spin}, although these calculations have not been worked out with the same accuracy as the Ceperley and Shumway work.

\section{Early Experiments on Exciton BEC in Cu$_2$O}

\subsection{Spectral analysis experiments}
\label{sfit}

The first experiments to show promise for exciton BEC in Cu$_2$O used excitation with a green laser which was absorbed within 5~$\mu$m of the surface \cite{wolfeprl1,wolfeprb}. This created a thin ``pancake'' of excitons which were not trapped, and could flow away from the surface, but on time scales of a few nanoseconds remained at high density at the point of creation.

\begin{figure}
\includegraphics[width=0.46\textwidth]{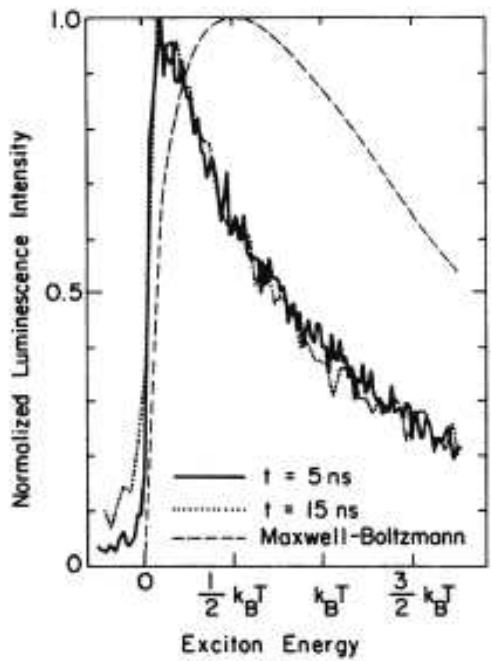}
\caption{Phonon-assisted luminescence spectrum showing the spatially-averaged energy distribution of orthoexcitons in Cu$_2$O for two different times as the density falls after the excitons are created by a short laser pulse. The fit temperature at 5 ns is approximately 60 K while the temperature at 15 ns is approximately 30 K. From Ref.~\protect\cite{wolfeprb}.}
\label{figscale}
\end{figure}

The phonon-assisted luminescence spectrum of the orthoexcitons (the upper, triplet state of the 1s excitons in Cu$_2$O) could be well fit by the Bose-Einstein distribution
\begin{equation}
I(E) \propto D(E)f(E) \propto E^{1/2}\frac{1}{e^{(E-\mu)/k_BT}-1},
\label{fit}
\end{equation}
with $\alpha \equiv |\mu/k_BT| \sim 0.1$ over a wide range of densities. In other words, the gas remained near the phase boundary for BEC but did not cross it. This was termed ``Bose saturation.'' Figure \ref{figscale} shows that this saturation corresponds to an energy scaling such that the shape of the photoluminescence remains the same. This scaling was found to apply over more than an order of magnitude of density variation.

The fits to the Bose-Einstein distribution imply an absolute density of the particles according to
\begin{equation}
N = \int_0^{\infty}\frac{1}{e^{(E-\mu)/k_BT}-1} D(E)dE.
\end{equation}
The densities from the fits were compared to data for the relative change of the density, found by taking the total luminescence intensity and dividing the measured exciton cloud volume at each point in time, and the variation in the density deduced from the fits agreed with the variation in density from the intensity data within a factor of two over a wide range of density \cite{wolfeprl1}. 

Later experiments looked at the paraexciton phonon-assisted luminescence under very similar experimental conditions and concluded that the paraexciton density exceeded the density for BEC predicted by the ideal gas theory \cite{linwolfe}. Line shape analysis of the paraexciton phonon-assisted emission was harder, because the emission intensity of this line is 500 times weaker than the orthoexciton line, and it lies near another, brighter emission line. The relative intensity of the paraexciton line seemed consistent with the conclusion that the paraexcitons exceeded the critical density. 

The first indication that this analysis could not be the whole story came from noting that the many-body collision broadening of the lines (discussed above) was very low, and not consistent with the densities implied by the BEC fits \cite{comments}. Although the exciton-exciton scattering cross section is not known to better than a factor of two, as discussed above, if the experiments were truly to have densities in the range of $10^{19}$ cm$^{-3}$, as implied by the fits of Eq.\,(\ref{fit}) to data like that shown in Fig.~\ref{figscale}, along with the observed spectral broadening of less than 0.1 meV, then these two observations together would require a scattering cross section with radius much less than the Bohr radius of the excitons. The crucial evidence came in comparing the exciton luminescence spectrum in the case of a cold (2 K) lattice, when the exciton distribution fit a Bose-Einstein distribution at 70 K, to the luminescence spectrum with a 70 K lattice under the same excitation conditions \cite{phcou}. The total number of photons emitted in each case was nearly the same, indicating that the densities were comparable, but the shapes of the spectra were quite different. The deviation from a Maxwell-Boltzmann distribution in the cold-lattice case could therefore not be from Bose statistics at high density if the same density gave a Maxwell-Boltzmann distribution when the system was closer to equilibrium with the lattice. This pointed to spatial inhomogeneity in a nonequilibrium system as the cause of the non-Maxwellian distribution. 

The main ambiguity of these experiments comes from the fact that the luminescence from the exciton gas must be integrated over at least one dimension, since the photons are collected by imaging a three-dimensional crystal in two dimensions. To fit the Bose-Einstein distribution to the data, some model of the spatial variation of the density and temperature must be used. This was recognized in the early work \cite{wolfeprb}, and it was found that nearly all models which assumed that the temperature and density of the orthoexcitons decreased monotonically in the direction away from the surface gave spatially integrated spectra that looked very similar to the simple homogeneous spectrum of Eq.~(\ref{fit}). (The difficulty of interpreting the three-dimensional data integrated over one dimension was one motivation for pursuing experiments with excitons, or exciton-polaritons, in two dimensions (for reviews see, e.g., Refs.~\cite{indirect} and \cite{polariton}.)

\begin{figure}
\includegraphics[width=0.66\textwidth]{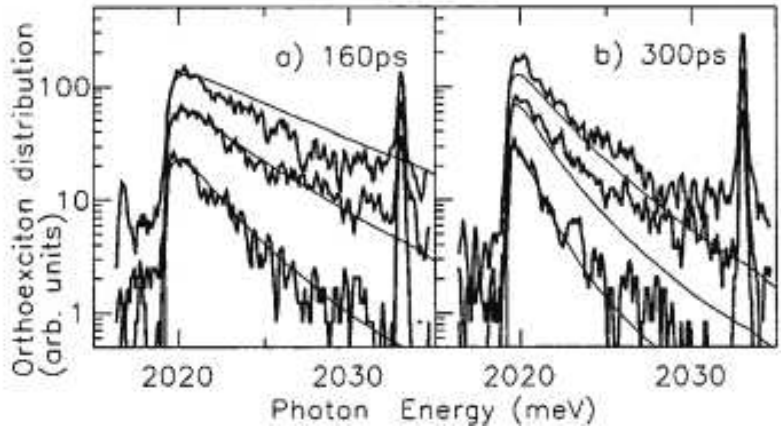}
\caption{Fits of a theoretical model for nonequilibrium flow of excitons to the phonon-assisted photoluminescence spectrum of Cu$_2$O with high-intensity surface excitation for three excitation powers under conditions similar to those of Fig.~\protect\ref{figscale}, at two different times after a short laser pulse. From Ref.~\protect\cite{wolfeohara}.}
\label{ohara}
\end{figure}

Later modeling \cite{wolfeohara} found that a reasonable fit to the spatially-integrated orthoexciton luminescence spectra could be obtained by assuming a large cloud of cold orthoexcitons far from the surface (see Fig.~\ref{ohara}). A simple analysis would seem to indicate that orthoexcitons do not travel very far within their lifetime due to down-conversion into paraexcitons \cite{orthodown,Denev,denev-conf}, and therefore one should not expect so many orthoexcitons away from the surface. However, the model of Ref.~\cite{wolfeohara} showed that picking a high enough rate for an exciton Auger recombination process could generate orthoexcitons at long distances from the surface due to up-conversion from paraexcitons.  

\begin{figure}
\includegraphics[width=0.9\textwidth]{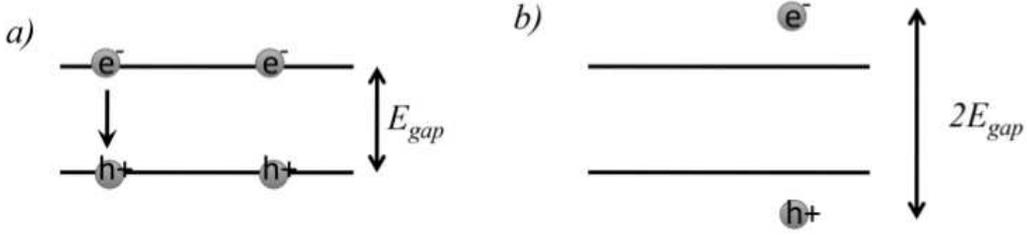}
\caption{The Auger recombination process. In (a), the exciton on the left recombines, giving its energy to the exciton on the right, which leads to the final state (b) of a single ionized exciton.}
\label{fig.auger}
\end{figure}

{\bf Auger recombination}. This brings up the Auger recombination process in Cu$_2$O, which was established to occur in Cu$_2$O in the 1980's \cite{trauer} and became an increasingly important topic of study in the 1990's. The Auger process occurs when two excitons collide, and one of them recombines, but instead of emitting a photon, the energy of the recombining exciton is given to ionizing the second exciton, as shown in Fig.~\ref{fig.auger}. The hot electron and hole thus produced can then lose energy by phonon emission or by collisions with other carriers, and finally form into an exciton again. It is assumed that the spins of the electron and hole are randomized in this process, so that the returning exciton can be either an orthoexciton or paraexciton. 

The Auger recombination process is density dependent, since it is proportional to the rate of excitons colliding with each other. The Auger rate is therefore parametrized by a constant typically called $A$, such that the rate of the process is
\begin{equation}
\frac{1}{\tau} = An,
\end{equation}
where $n$ is the density. For a single population with an intrinsic radiation lifetime $\tau_r$, this gives the rate equation
\begin{equation}
\frac{\partial n}{\partial t} = G(t) -\frac{n}{\tau_r} - \frac{1}{2}An^2,
\end{equation}
where $G(t)$ is the generation rate of the excitons. A factor of $1/2$ occurs because half the excitons are returned to the population after ionization. For a two-population system such as the orthoexcitons and paraexcitons in Cu$_2$O, we must write two coupled rate equations,
\begin{eqnarray}
\frac{\partial n_o}{\partial t} &=& G_o(t) -\frac{n_o}{\tau_{or}} - An_o^2 - An_on_p +\frac{1}{2}\alpha A (n_o+n_p)^2\nonumber\\
\frac{\partial n_p}{\partial t} &=& G_p(t) -\frac{n_p}{\tau_{pr}} - An_p^2 - An_on_p + \frac{1}{2}(1-\alpha)A(n_o+n_p)^2, \label{simplerates}
\end{eqnarray}
to take into account the collisions between different species. The last term in each equation gives the effect of the return of the ionized electrons and holes; the factor $\alpha$ gives the fraction of these which return to each species. It is typically assumed that $\alpha = 3/4$, since the orthoexciton population has three degenerate states while the paraexciton is a singlet. An additional modification of these equations can be made to take into account direct conversion of one species into the other by phonon emission and absorption \cite{orthodown,Denev,denev-conf}. Most generally, the constant $A$ can be different for collisions of different types of excitons, for example due to the effects of indistinguishability on the different spin states \cite{KB}. Writing down down three parameters, $A_{oo}$, $A_{op}$, and $A_{pp}$, the rate equations (\ref{simplerates}) become
\begin{eqnarray}
\frac{\partial n_o}{\partial t} &=& G_o(t) -\frac{n_o}{\tau_{or}} - A_{oo}n_o^2 - A_{op}n_on_p +\frac{1}{2}\alpha (A_{oo}n_o^2+2A_{op}n_on_p + A_{pp}n_p^2)\nonumber\\
\frac{\partial n_p}{\partial t} &=& G_p(t) -\frac{n_p}{\tau_{pr}} - A_{pp}n_p^2 - A_{op}n_on_p +\frac{1}{2}(1-\alpha) (A_{oo}n_o^2+2A_{op}n_on_p + A_{pp}n_p^2). 
\end{eqnarray}

The Auger process 1) reduces the lifetime of the excitons dramatically at high density, from the radiative lifetime of milliseconds down to less than a nanosecond, 2) heats the exciton gas due to the gap energy given to the ionized electron and hole, which then interact with the rest of the system, and 3) gives a population of orthoexcitons in excess of that expected from the Boltzmann occupation factor $e^{-\Delta/k_BT}$, where $\Delta$ is the paraexciton-orthoexciton splitting. Such an imbalance implies that the system is not in full equilibrium, which can occur when the lifetime due to Auger annihilation of the paraexcitons is shorter than the ortho-to-para conversion time. 

Throughout the 1990's and early 2000's, several groups tried to measure the rate of Auger recombination in Cu$_2$O. This work is reviewed in Section~\ref{augerrev}. The measurements are made difficult by the fact that there is also a two-body conversion process by which two orthoexcitons can collide and turn into two paraexcitons. Since this process has the same density dependence as the Auger process it can be difficult to distinguish the two processes in many experiments. In recent experiments, the group of Kuwata-Gonokami used infrared absorption measurements to excite the 1s-2p exciton transition \cite{gono-lyman1,gono-lyman2,gono-lyman3}, which is analogous to the 1s-2p transition in alkali atoms. Since this transition only occurs when the excitons exist, the absolute absorption cross section of the transition gives an absolute measurement of the number of both species of excitons. Therefore the spin flip process, which conserves total number of excitons, could be distinguished from the Auger process, which annihilates excitons. 

Although there was a range of values found for the Auger rate, the general consensus arising from the Auger measurements has been that the density of the excitons could not be as high as predicted by the fits of the surface-excitation data to the Bose-Einstein distribution, because the lifetime would be too short at those densities. Since, as discussed above, there was also inconsistency in the broadening of the spectral lines and the absolute number of photons coming from the phonon-assisted luminescence, it was generally concluded by the end of the 1990s that the early experiments did not show Bose-Einstein condensation. The Bose saturation effect, which corresponds to an invariance of the luminescence spectrum as temperature is scaled (see Fig.~\ref{figscale}), is therefore surprising, because there are a number of details which go into the classical numerical model of which have to be fortuitously related to give the nearly invariant spectrum which is observed; without these coincidences the spectrum would have different shapes at different times.  The model of Ref.~\cite{wolfeohara} used choices of the physical parameters which were reasonable based on the experimental data, but even with the optimal choice of realistic parameters, the fit of the model to the data is qualitatively similar in shape but not especially good (see Fig.~\ref{ohara}). The saturation effect corresponds to a scaling law, which we discuss below, in Section~\ref{sat}.



\subsection{Transport experiments}

Also during the 1990's, another set of experiments was performed that focused on transport of the excitons rather than spectral signatures \cite{fortin1,fortin2,fortin3,fortin4}. The experiments were done as follows: a long-wavelength (red) laser with photon energy nearly resonant with the exciton energy was used to flood a whole three-dimensional crystal of Cu$_2$O with a low-density exciton gas. Then a very intense pulse of green light was used to excite one surface of the crystal, creating a very dense exciton cloud at the surface. Excitons were then detected on the opposite side of the crystal, in some cases over a millimeter distant, by a bimetal detector that ionized the excitons into free electrons and holes which then produced an external current. The signal would only occur if excitons migrated across the crystal into the detector.

The experiments showed that the exciton signal on the back side of the crystal jumped up in a sharp pulse when the green laser pulse intensity exceeded a critical threshold. The arrival time of the pulse corresponded to propagation at the speed of sound across the crystal. This was interpreted by the authors as superfluid motion of the excitons across the crystal when the excitons exceeded a critical density threshold. The high density pulse of excitons created by the green pulse was viewed as being amplified by the large population of cold excitons created by the resonant laser. 

There were soon alternative interpretations of these experiments.  Ref.~\cite{kav-pulse} pointed out that the width of the pulses was orders of magnitude longer than the expected coherence length, and that a classical soliton model could reproduce many of the effects. Tikhodeev and coworkers \cite{tikh1,tikh2,tikh3,tikh4} pointed out  that the intense green laser pulse also created many hot phonons which could blow as a nonequilibrium ``phonon wind'' away from the surface, pushing excitons at the speed of sound, and reasonable models of the phonon wind could explain the exciton propagation. The phonon wind was well established (e.g., Ref.~\cite{phonwind,phonwindwolfe}) in the 1980's as an effect in surface-excited solids; a flux of phonons streaming away from the surface can push excitons and electron-hole liquid for hundreds of microns. One experimental result in support of this interpretation is that a second pulse was seen at the detector at much later times, corresponding to three passes across the crystal at the speed of sound, i.e., hitting the back surface, bouncing off, and coming back one round trip later \cite{andreprivate}.  This strongly supports the view that a sound pulse was created by the green laser pulse, and this sound pulse propagated across the crystal and reflected from the surfaces, sweeping excitons from the cold exciton gas as it traveled. 

As with the spectral signature experiments, it was widely concluded by the end of the 1990's that these experiments with surface excitation also did not show Bose-Einstein condensation. Both of these experiments used very intense surface excitation in order to get very high exciton density, but both had the same drawback that the exciton gas thus created was very far from equilibrium and unconstrained in space, so that flow out of the excitation region had to be modeled, including the heat flow and phonon wind. Because of this, attention returned to trapping the excitons using stress.


\subsection{A Way Forward}

It became clear by the first ICSCE conference in 2003 \cite{icsce1} that the Auger effect played a dominant role in preventing EBEC in Cu$_2$O by shortening the lifetime of the excitons. To avoid this, the density of the excitons can be dropped, with a corresponding drop in the temperature, since $T_c$ of the BEC transition is proportional to the density to the 2/3 power in three dimensions. At first, one might think that dropping the density would not help, because the collision rate between the particles has the same dependence on density as the Auger rate, and therefore the ratio of the Auger lifetime to the collision time will not change. Excitons in a semiconductor have another means of thermalizing besides collisions, however, which is exciton-phonon scattering. The time scale for exciton-phonon equilibration is hundreds of picoseconds to nanoseconds \cite{snokebc,ivanov_longthermtime} at low temperature, so that when the gas has low density, the time scale for thermalization via phonons can be much shorter than the lifetime due to Auger recombination. 

One issue in some solid-state systems in going to lower temperature and density is that the excitons can become localized in minima of the disorder (e.g., \cite{zoltan_local}), so that they do not move and do not thermalize. Many Cu$_2$O samples have extremely good quality, however, so that disorder is not a limiting factor. Another issue is that the polariton effect, i.e., mixing of the exciton and photon states, becomes important for excitons at low momentum, i.e., low temperature. There is a polariton effect in Cu$_2$O for both the orthoexciton species and for the paraexciton species in stressed crystals; although the direct recombination process for paraexcitons is forbidden by symmetry in unstressed crystals, if stress is used to create a trap for the excitons, the paraexciton coupling to the photons becomes comparable to that of the orthoexcitons \cite{trauer}. It is possible, however, to go to temperatures so low that the excitons have momentum which is below the mixing region of the photon and exciton states. 

The above considerations imply that the best temperatures to see BEC effects in Cu$_2$O are in the sub-Kelvin range of tens to hundreds of mK. This removes one of the original motivations for studying exciton condensates, which is to see high-temperature condensation; the fact that the excitons in Cu$_2$O are stable up to room temperature turns out not to help in getting them to condense at room temperature. One can still hope to see a true exciton BEC in a three-dimensional system, however.  Two experimental groups have pursued EBEC experiments of excitons in traps in Cu$_2$O at milliKelvin temperatures in the past five years; these are reviewed in Section~\ref{recent}.

\section{Review of the Auger process}
\label{augerrev}

\vskip2pc

As discussed above, one of the most important of the mechanisms which determine the lifetime 
of excitons in Cu$_2$O is believed to be a nonradiative
two-body decay process, as evidenced by numerous experiments which have shown that the exciton lifetime is density dependent. 
This ``Auger" process is a well-known decay mechanism in the
physics of semiconductors. This process is also present in 
electron-hole plasmas, in which the recombination of an electron 
and a hole excites either an electron high in the conduction band 
or a hole deep in the valence band. This process, which occurs in many semiconductors, has long been studied 
both theoretically \cite{BL58,ridley,haug-auger,das-auger} and experimentally \cite{GW81,yosh}. A 
similar decay process has also been observed in experiments with cold 
metastable He atoms, where the term ``Penning collisions" is used 
\cite{Penning}. It also occurs with Frenkel excitons in organic materials, where it can be quite strong \cite{org1,org2}. There may, in fact, be a general principle that excitons with large binding energy, and therefore are Frenkel-like, intrinsically will have strong Auger recombination. This is seen in the phonon-assisted Auger theory discussed below, which as seen in Equation~(\ref{augercalc}), gives a rate proportional to $1/a^2$, where $a$ is the exciton Bohr radius. 

In Cu$_2$O, the exact nature of the Auger process 
is still an open question and is under current investigation. As discussed above, the general picture is that in such a process two excitons collide, the one  recombines, transferring its energy to the other, which ionizes. The 
electron and the hole which result from the ionization process then 
form an exciton, after a fraction of their energy is transferred to 
the lattice, via phonons. The final states of the ionized electron and hole, however, are not entirely clear. 

Before we review the main results on this process, we start mention some reasons which make the study 
of this process in Cu$_2$O difficult: 

(1) The experiments which have measured this process cover a rather wide range of exciton temperatures $T$ and exciton densities 
$n$. The Auger rate is likely to depend on the temperature, but the temperature of the excitons may not be the same as the lattice temperature in all cases, making it harder to determine the dependence on temperature accurately. 

(2) In excitons in Cu$_2$O there are many different processes which 
take place at the same time (radiative recombination, conversion 
between the orthoexcitons and the paraexcitons, expansion, etc.) These 
processes obscure the measurement of the Auger decay process alone, 
since there is strong evidence that the rate of some of the 
above mechanisms is comparable to the rate of the Auger process. 

In particular, the interconversion process of the total angular-momentum 
triplet state orthoexcitons into the total angular-momentum singlet 
state paraexcitons is of great importance. It is believed that it 
takes place via two mechanisms, namely a phonon-assisted process 
\cite{CW, orthodown, denev-conf, Keith, WolfeJang} and a collisional spin-exchange 
process \cite{KM}. At low temperature and high density, the collisional ortho-para conversion process in many ways mimics the Auger collisional recombination process, making analysis much more difficult. 

As a result of the many processes which all take place at the same 
time, one has to write down rate equations, which then give the Auger 
rate (as well as the rate of the other processes) from some fitting. 
Therefore, to a large extent, the determination of the Auger rate is 
model-dependent, and this complicates the problem. In many cases, the models must include parameters for spatial diffusion of the excitons in order to estimate the volume of the exciton cloud. 

In some regimes of parameters, different processes can be isolated. For example, experiments can be done in a very low density regime in which collisions between excitons are negligible. In this case, the phonon-assisted conversion between orthoexcitons and paraexcitons can be determined very accurately \cite{orthodown,Denev}. It is also possible to establish the intrinsic lifetime of the excitons by fitting the decay of the luminescence at very low densities, when the Auger recombination rate is negligible. In addition, at temperatures which are high compared to the energy splitting between the ortho and para states, chemical equilibrium between the ortho and para populations can be assumed, so that all ortho-para conversion processes drop out of the rate equations. This fact has been used to extract the total Auger rate of the exciton population in the high-temperature regime \cite{yingmei}, although these results still relied on a model for the exciton cloud volume.

(3) The determination of the exciton density has proven to 
be a rather difficult task. There are basically three approaches that 
one may follow to extract the density. The first is the spectroscopic approach, in which some 
specific phonon-assisted recombination line is fitted to a 
Bose-Einstein distribution and from that the density and the 
temperature are extracted \cite{wolfeprl1}, as discussed in Section~\ref{sfit} above. As discussed above, the estimates of density based on spectroscopic fits to an equilibrium homogeneous gas can strongly overestimate the density. The second method of extracting the exciton density is based on 
estimating the volume of the exciton gas and the number of excitons  \cite{phcou,wolfeohara,OHara}. 
The third method relies on the absorption spectrum that corresponds to 
the 1s to 2p radiative excitonic transition. This method does not 
depend on the strength of the recombination lines and is thus 
advantageous in that respect. In addition, the absorption spectrum of this process is very sensitive to the 
degeneracy of the excitons and can serve as a clear and indisputable 
probe for the transition to a Bose-Einstein condensed phase \cite{KJ}. 

(4) The theoretically-calculated decay rate involves several parameters
with experimental uncertainty in their numerical value. As a result, the theoretical estimate of the 
decay rate is also uncertain.

The importance of the Auger process in excitons in Cu$_2$O has been 
revealed in several experiments over the last three decades 
\cite{naka-njp,WolfeJang, HMBa,HMBb, trauer, SW90b, yingmei,OHara, Warren, wolfeohara, 
Denev, Jolk, gono-lyman2, Jang, gono-lyman3}. The decay rate
of the excitons shows an approximate proportionality 
to their density, which is an indication of a two-body collision 
process. The Auger process is also confirmed by the presence of orthoexcitons 
in stress-generated traps for times much longer than the 
orthoexciton lifetime \cite{trauer, SW90b}; as discussed above, high energy electrons and holes produced by the 
Auger ionization of para excitons lead to re-formation of excitons in 
essentially random internal angular momentum states, and thus to orthoexcitons in excess of the amount expected from Boltzmann occupation. 

The difficulties discussed above have led to a wide scatter in experimental estimates of the Auger decay rate. The decay 
rate has been extracted to be negligible \cite{Jolk} and also non-negligible, and when non-negligible, to be independent of the temperature \cite{gono-lyman3}, to increase 
linearly with the exciton temperature \cite{yingmei}, and to be 
inversely proportional to the exciton temperature \cite{WolfeJang}. 
The stress dependence of the Auger decay rate is also of importance 
\cite{Denev}. The process is sensitive to the various symmetries of 
the crystal, which impose certain selection rules/constraints 
on the matrix elements and on the decay rate, while stress tends to 
lift these constraints. It has also been argued \cite{Jang} that 
excitons form biexcitons, which then rapidly decay via the Auger 
process; there is no direct experimental evidence for this process, however, to our knowledge. In addition, the theoretical calculation 
presented in Ref.~\cite{Jang} ignores the orthogonality of the bands 
of the crystal. (According to the authors of this study, the Auger 
mechanism is thus associated with breaking the band symmetries, e.g. by impurities, and therefore differences of impurity concentrations could explain the different Auger rates measured). 

\subsection{Review of Auger theory, constrained by experiments}
\label{auger}

\begin{figure}
\includegraphics[width=0.35\textwidth]{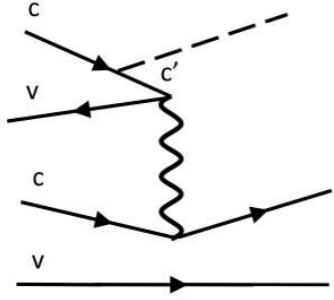}
\caption{A phonon-assisted Auger process in which two excitons (electron-hole pairs) collide, and one recombines, transferring its energy to another exciton via a Coulomb interaction. A phonon emission causes an electron to transition to a virtual intermediate state. There are four nonequivalent diagrams for the phonon-assisted Auger process, since either an electron or hole can emit a phonon, and either an electron or hole can receive the momentum from the recombining exciton.}
\label{fig.dia}
\end{figure}
Turning to the theoretical study of the Auger decay rate, according 
to Ref.\,\cite{KB} this process may be either ``direct", or phonon 
assisted. In the phonon-assisted process there is a phonon involved, as shown in the diagram in Fig.~\ref{fig.dia}. While the participation of a phonon suppresses the rate of this process, on the other hand, it lifts the 
symmetry constraints set by the band structure of Cu$_2$O. We should 
stress here that there is an intimate connection between the radiative 
and the Auger decay mechanisms: the radiative lifetime of excitons is 
rather long as the dipole matrix element between the conduction and the 
valence bands (which have the same parity) vanishes. Phonon-assisted 
mechanisms provide an alternative way for the excitons to decay 
radiatively. As in the Auger process, the exciton-phonon interaction 
suppresses the rate of this process, but on the other hand it makes the 
process dipole allowed. As a result, the phonon-assisted processes are 
actually the dominant radiative decay mechanisms. Rather similar 
results also hold for the direct and the phonon-assisted Auger 
processes: the theoretical calculations indicate that the dominant Auger process is the 
phonon-mediated process. Although in this case one also needs a phonon, which makes it less likely, on the other hand, there is no symmetry constraint, which enhances the rate of this process. 

The dependence of the matrix element for the Auger 
collisions on the momentum exchange between the colliding excitons 
is of major importance \cite{KB}. If this is constant, then its only 
temperature dependence comes from phonon emission \cite{yingmei}. If the 
matrix element is proportional to the momentum exchange, the Auger 
decay rate is proportional to the exciton temperature as well as to the 
density of the excitons. 


\begin{figure}
\includegraphics[width=0.7\textwidth]{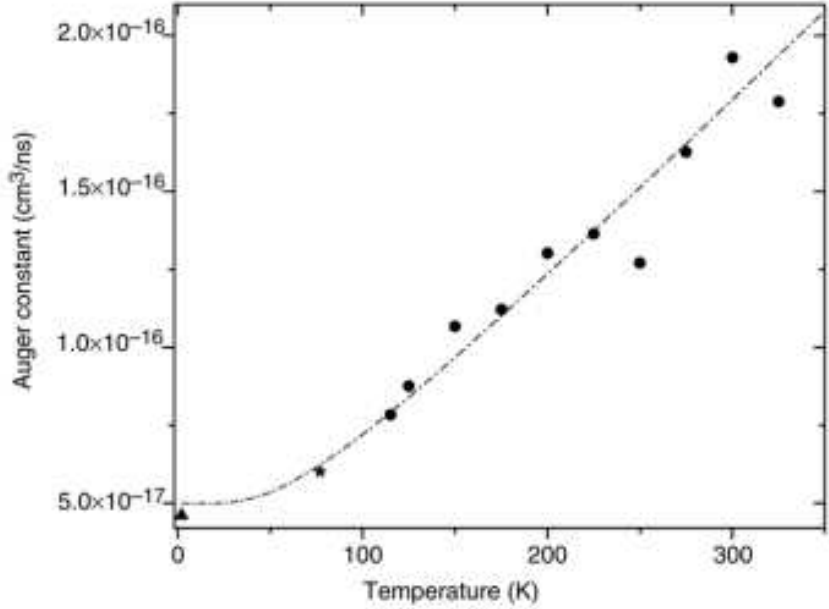}
\caption{The measured Auger rate from several measurements. Circles: measurements from Ref.~\protect\cite{yingmei}. Triangle: measurement from Ref.~\protect\cite{Denev}.  Star: measurement from Ref.~\protect\cite{Warren}. Dashed-dotted line: phonon-assisted Auger recombination theory from Refs.~\protect\cite{KB} and \protect\cite{yingmei}, discussed in the text.}
\label{yingmeifig}
\end{figure}
Based on Ref.~\cite{KB}, Ref.~\cite{yingmei} evaluated the Auger rate  
$\tau_{\rm Auger}^{-1} = A n_{o}$ for orthoexciton-orthoexciton 
collisions. The formulas given in Ref.~\cite{yingmei} contained some errata; the correct formula for the Auger rate considered in Ref.~\cite{KB}, integrated over all $k$-vectors, is (in MKS units)
\begin{eqnarray}
 \frac{1}{\tau_{\rm Auger}} &=& n_o \frac{128\hbar}{\pi^3}\left( \frac{m_h^{3/2}}{m_e^{1/2}m} \right)\left(\frac{2\langle p\rangle^2}{m E_{c'v}}\right)\left(\frac{e^2}{\epsilon a}\right)^2 \frac{D^2}{\rho  E_{opt}}\frac{1}{E_{c'v}E_{cc'}^2}
\left(\int_0^b \frac{x^4\sqrt{b^2-x^2} dx}{(1+x^2)^6}\right) \nonumber\\
&& \times \left(1+\frac{1}{e^{E_{opt}/k_BT}-1}\right),  \label{augercalc}
\end{eqnarray}
with
\begin{equation}
b^2 = \frac{2m_ea^2E_{c'v}}{\hbar^2},
\end{equation}
where $m_e$ and $m_h$ are the effective mass of the electron and hole, respectively, $m$ is the standard electron mass, $\epsilon$ is the dielectric permittivity of the medium, $\langle p\rangle$ is the dipole moment for optical inter band transitions, $\rho$ is the crystal mass density, $D$ is the deformation potential for excitons interacting with optical phonons with energy $E_{opt}$,  $a$ is the exciton Bohr radius, and $E_{c'v}$ and $E_{cc'}$ are the energy differences between a higher-lying conduction band $c'$ (which is assumed to be involved in virtual transitions) and the valence and conduction bands $v$ and $c$, respectively.  All of the numbers in this formula are known from other experiments. The optical phonon deformation potential is known from time-resolved experiments \cite{snokebc} to be on the order of a few times $10^7$ eV/cm; the oscillator strength $f = 2\langle p\rangle^2/m E_{c'v}$ can be assumed to be of order unity \cite{snokebook}; the interband energies are known and are all on the order of an eV; the optical phonon energies have been measured accurately and are on the order of 10-20 meV; the exciton Bohr radius is known to be around $5$~\AA. (See Ref.~\cite{yingmei} for the experimental references for these numbers.)  These numbers give $A$ on the order of $5\times 10^{-18}$ cm$^3$/ns in the low-temperature limit.  In other words, an exciton gas with density $10^{17}$ cm$^{-3}$ will have a lifetime of around 2 ns due to Auger recombination, much shorter than the typical radiative recombination times in the low-density limit.

Ref.~\cite{yingmei} found a best fit, shown in Fig.~\ref{yingmeifig}, of 
\begin{eqnarray}
A   &\approx& 3.698 \times 10^{-17}  \left( 1 + \frac 
   2 {e^{160/T} - 1} \right) + 1.298 \times 10^{-17} \left(
   1 + \frac 2 {e^{217/T} - 1} \right),
\label{the}
\end{eqnarray}
for $A$ in units of cm$^3$/nsec and the temperature $T$ in Kelvin.
The two exponential factors 
come from the emission of the optical phonons $\Gamma_{12}^-$ and 
$\Gamma_{15}^-$, with energies 13.8 meV and 18.7 meV, respectively. 
The temperature dependence is in good agreement with the experimental 
measurements for a wide range of temperature between 
2 K and 235 K with essentially two fitting parameters, namely the oscillator strength $f$ times the deformation potential $D$ for each phonon-assisted transition; the fit value of this product is within a factor of 3 of the value estimated above. According to Eq.\,(\ref{the}),  for $T$ roughly up to 77 K the rate is independent of $T$, while for 
higher temperatures the rate increases linearly with $T$ (which is 
obvious if one expands the exponentials). 

References \cite{OHara} and \cite{Warren} estimated the exciton 
number and the volume of the gas at relatively low temperatures (and 
thus did not get the density spectroscopically). The values of the Auger 
constant $A$ that they found using this method were $7 \times 
10^{-17}$ cm$^3$/ns and $6 \times 10^{-17}$ cm$^3$/ns, respectively, which is in decent 
agreement with the result of Eq.\,(\ref{augercalc}) and with the experimental 
results of Ref.~\cite{yingmei} at low temperature. Reference \cite{naka-njp} also recently measured the orthoexciton-orthoexciton Auger rate to be $5\times 10^{-17}$ cm$^3$/ns at low temperature, very close to the low-temperature value of Ref.~\cite{yingmei}.

For temperatures above 100 K, it is worth a close look to compare the results of Refs.~\cite{WolfeJang} and \cite{yingmei}. As noted above, the values for the Auger constant of the two groups at low temperature fall within the same range, with about a 30\% deviation, which is not surprising given the various experimental uncertainties. However, in the range 100-200 K, the two experiments showed opposite trends with temperature: the Auger rate was found to rise by about a factor of two in Ref.~\cite{yingmei} and to fall by a little less than a factor of two in Ref.~\cite{WolfeJang}. The two experiments used similar methods, with pulsed near-resonant dye laser excitation. Even apart from any modeling, the raw data of Ref.~\cite{yingmei} clearly showed that the excitons had shorter lifetime at high temperature, while the raw data of Ref.~\cite{WolfeJang} shows the opposite trend. 

The primary difference between the experiments was that a much tighter focus was used in the experiments of Ref.~\cite{WolfeJang}, about 15 $\mu$m radius, as compared to 120 $\mu$m radius in Ref.~\cite{yingmei}. This gave higher densities and much greater range of lifetime, but may also have had other effects. One possibility is that the local lattice temperature may have gone well above the background bath temperature due to the concentration of energy, leading to a local ``hot spot'' (cf., e.g., Refs.~\cite{hotspot1,hotspot2}.) The exciton temperature may also have gone well above the lattice temperature, if the Auger rate was fast compared to the phonon emission rate.  Taking the Auger rate as increasing with temperature as discussed above, if a local exciton cloud is hotter than the rest of the crystal, a higher Auger rate will occur.  To actually have a decrease of the Auger rate with increasing lattice temperature, however, would require that the local temperature excursion was greater at low bath temperatures. It is also possible that in the experiments of Ref.~\cite{yingmei}, at the highest temperatures there was also an excursion of the temperature above the lattice temperature, because the absorption length of the input laser shortened to approach 20 $\mu$m (although the lateral size still gave much larger volume for the spot size than the experiments of Ref.~\cite{WolfeJang}). 

Alternatively, or possibly in conjunction with a local hot spot, the rapid expansion of the exciton cloud seen in Ref.~\cite{WolfeJang} may have played a role. In the experiments of Ref.~\cite{yingmei}, volume expansion was a minor factor because the initial size of the cloud was large. With the smaller initial spot size in Ref.~\cite{WolfeJang}, the expansion of the cloud played a bigger role. The authors of Ref.~\cite{WolfeJang} carefully estimated the volume as a function of time $V(t)$ based on a model of diffusion to get the density $n(t) = N(t)/V(t)$, but the relative uncertainty of the volume will be larger with a smaller spot size.  

At present, Refs.~\cite{gono-lyman3} and \cite{Jolk} are the outliers. Ref.~\cite{gono-lyman3} reported a value for the Auger rate of approximately $4 \times 10^{-16}$ at low temperature, and Ref.~\cite{Jolk} reporting a much lower, indeed negligible, rate, of the order of 10$^{-23}$ cm$^3$/ns.  The majority of other experimental work \cite{yingmei,WolfeJang,naka-njp} gives Auger rates in the range of 10$^{-17}$ to 10$^{-16}$ cm$^{3}$/ns. Ref.~\cite{gono-lyman3} reported uncertainty of a factor of three, which would allow a value in this range by taking the low side of the uncertainty range. 

In the case of the work by the group of Gonokami, a multi-step process of analysis and measurements was used to derive the Auger rate. In the first step, the absolute number of excitons is established by far infrared Lyman absorption spectroscopy of the 1s-2p, 1s-3p, etc. transitions for both orthoexcitons and paraexcitons \cite{gono-lyman1,gono-lyman2,gono-lyman3}. The reason that the Lyman spectroscopy can establish the absolute number of excitons while band-to-band absorption cannot is that the absolute cross section for the Lyman series can be very accurately calculated from first principles for intraband transitions, while the band-to-band oscillator strength cannot. 

Having this method to establish the absolute number of both ortho and paraexcitons, the number must be converted to a density by an estimate of the volume of the exciton cloud. In a stress-generated harmonic trap, this can be done by imaging the exciton cloud from two sides of the crystal, but typically the number of excitons in such a trap is too few to generate a reasonable Lyman absorption measurement. Therefore a large exciton cloud was generated in a bulk crystal, without a trap, and a model for the exciton diffusion was used to estimate their volume. This modeling is the greatest source of uncertainty in the measurements, since various effects such as phonon wind, discussed above, may also affect the exciton motion. The authors of Ref.~\cite{gono-lyman3} used the diffusion constant for paraexcitons reported in Ref.~\cite{trauer-low}, which was very high. This was reasonable, given that their sample was cut from the same natural sample as that used in Ref.~\cite{trauer-low}, but was not checked directly. If the diffusion constant was estimated too high, then the volume would be estimated as much greater, and therefore the density much lower, which could give systematically high Auger rates.

The work of Ref.~\cite{Jolk} gave the opposite result of very low Auger rate for both paraexcitons and orthoexcitons. In this experiment, the absolute density of the excitons was estimated by using the effect of screening of excitons at high density on the absorption spectrum \cite{jolk-screen}, based on a theory of the Mott transition. Their theory of the screening led them to estimate initial densities on the order of 10$^{19}$ cm$^{-3}$, and under these conditions they saw no evidence of an Auger effect. The theory of the excitonic Mott transition is hotly debated and quite complex, however \cite{exion,manzkenew1,manzkenew2,craw}, and therefore it is difficult to find agreement on a universal theory for the density at which the exciton absorption will be screened out. 

The above experiments primarily were sensitive to the ortho-ortho Auger recombination rate. The question of the value of the Auger collisions between paraexcitons is even more more subtle.  Reference \cite{naka-njp} has provided evidence 
for a very low paraexciton-paraexciton Auger rate, on the order of $10^{-18}$ cm$^3$/ns. Earlier
experiments of Refs.~\cite{Denev} in stressed crystals also gave a para-para Auger rate about two orders of magnitude less than the ortho-ortho rate.  However, other experiments suggest that Auger 
collisions are also possible among paraexcitons \cite{OHara}. Reference \cite{gono-lyman3} used a single Auger rate constant for all collisions in their model, so that their fit was not sensitive to a difference between the para-para, ortho-para, and ortho-ortho rates. 

Theoretically the para-para Auger collisions should be 
negligible \cite{KB} at zero stress (the effect of stress is discussed 
below). The reason for this is the band structure of Cu$_2$O, which 
makes the direct Auger process for para-para collisions forbidden, 
while the phonon-assisted process is highly suppressed \cite{KB}. This 
is due to the fact that the phonon-assisted Auger process for 
para-para Auger collisions involves a virtual transition to a deep valence band, as opposed 
to the ortho-ortho collisions, for which this process involves a 
conduction band which is close in energy to the lowest conduction band.

\subsection{Orthoexciton to paraexciton conversion process and
``saturation" of the orthoexciton gas}
\label{sat}

As discussed above, at very low densities, when the two-body processes are negligible,
it has been observed that orthoexcitons convert into paraexcitons
through a phonon-assisted process \cite{CW, orthodown, Keith, WolfeJang}. According to Ref.\,\cite{orthodown} 
its rate scales as $T^{3/2}$. More recently Refs.~\cite{Keith} and \cite{WolfeJang} have reported that its rate is constant at very low 
temperatures and increases linearly with the temperature at higher 
temperatures. 

In addition to the phonon-assisted process, orthoexcitons convert into
paraexcitons via a spin-exchange process, where two orthoexcitons 
collide, they exchange their electrons or holes, resulting into two 
paraexcitons in the final state. This process was studied 
theoretically in Ref.\,\cite{KM}, while Ref.\,\cite{gono-lyman2} has 
shown experimental evidence for it. The decay rate of this process is 
proportional to the orthoexciton density $n_{o}$. 

A simple argument has been proposed at various times over the years (see, e.g., Refs.~\cite{KBW} and \cite{GK}) to explain the saturation effect of orthoexcitons, discussed in Section~\ref{sfit}, which corresponds to a power law $n_{o}  \propto T^{3/2}$. 
Recall that the exciton gas, which in general has a higher temperature than the lattice temperature, 
exchanges phonons with the lattice and thus loses energy, while on 
the same time it is heated by the Auger process and the ortho-to-para 
conversion. From simple deformation potential theory, the rate of loss of 
energy is proportional to $T^{3/2}$ (see Ref.~\cite{snokebook}, Section 5.1.4). On the other hand, 
both the Auger process and the collisional conversion process of orthoexcitons 
to paraexcitons heat the gas. In both cases the rate of energy gain 
of the gas is $b n_o$, where $n_o$ is the orthoexciton density. As a 
result, we can write a rate balance equation for the entropy of the
orthoexcitons divided by the orthoexciton number $S_o/N_o$
\begin{eqnarray}
  T \frac d {d t} \left( \frac {S_{o}} {N_o} 
  \right) \approx - a T^{3/2} + b n_o,
\end{eqnarray} 
which implies that the orthoexcitons move along adiabats, where 
$n_o \propto T^{3/2}$, in agreement with the experimental
data \cite{SW90b}. This above argument applies both in the classical and in the degenerate 
regime \cite{GK}.

\begin{figure}
\includegraphics[width=0.45\textwidth]{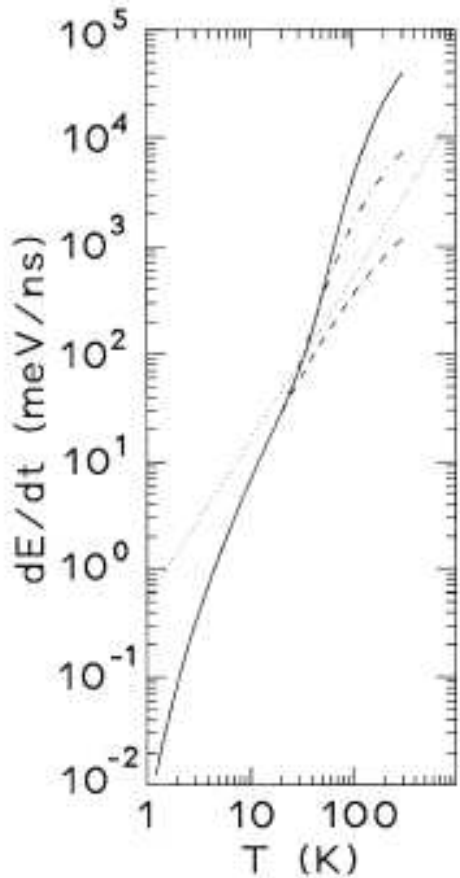}
\caption{Energy loss rate for orthoexcitons in Cu$_2$O as a function of exciton temperature, for a cold lattice. Dashed line: rate when only acoustic phonon emission is included. Dotted dashed line: rate when acoustic-phonon and optical single-phonon emission are included. Solid line: rate when acoustic, optical single-phonon, and optical two-phonon emission are included. Dotted line: the $T^{3/2}$ power law.  From Ref.~\protect\cite{snokebc}.}
\label{heat}
\end{figure}

A problem with this argument, however, is that the rate of heat loss for orthoexcitons in Cu$_2$O never exactly follows the $T^{3/2}$ law predicted by elementary deformation potential theory. There are both high-temperature corrections, due to optical phonon emission and high-frequency corrections of the exciton wave function, and low-temperature corrections due to momentum-conservation limitations \cite{trauer-low}, that give deviations from this law. The rate of phonon emission in Cu$_2$O has been measured carefully and reported in Ref.~\cite{snokebc}; a summary plot is shown in Fig.~\ref{heat}.

Ref.~\cite{wolfeohara} also addressed the saturation effect, using a classical rate equation model of with only acoustic phonon emission for energy relaxation and including the Auger process. They did not obtain a $n \propto T^{3/2}$ power law but did find a general trend of temperature increasing with density comparable to the experimental results. 

\subsection{The effect of stress}

As we mentioned earlier, the Auger decay process is strongly 
influenced by stress. In stressed crystals the selection rules which 
determine the Auger decay rate are no longer valid, and as a result 
stress enhances the rate of this process. 

The most important issue in the case of stressed crystals is the 
effect of stress on the paraexciton-paraexciton Auger collisions. As
discussed above, in non-stressed crystals it is an open question whether the para-para Auger process is allowed,
but it is certain that under stress it is allowed. 
An analogous situation is the radiative decay of paraexcitons, where 
stress makes the direct radiative recombination process of 
paraexcitons allowed, and it also enhances the phonon-assisted 
recombination lines. More specifically, uniaxial stress mixes the 
$\Gamma_7^+$ and $\Gamma_8^+$ valence bands with the result that the 
para-exciton energy increases for small values of the applied stress 
but decreases for higher values, and the direct radiative recombination 
increases quadratically with increasing stress \cite{parastress,pstr1,pstr2,pstr3}.

Indeed, the experiment in Ref.~\cite{trauer} shows clearly that 
para-para Auger collisions are allowed in stressed crystals. Uniaxial 
stress makes the direct Auger process allowed as well as the 
phonon-assisted process for paraexcitons. In both cases the orientation of the 
uniaxial stress is very important, since it determines the selection 
rules in the deformed crystal \cite{parastress,pstr1,pstr2,pstr3}. These processes are 
responsible for the presence of orthoexcitons in the stress-generated trap 
Ref.~\cite{trauer}. The
orthoexcitons are generated from free electrons and holes from 
Auger-ionized paraexcitons; the free electrons and holes are assumed to enter random angular momentum states. 

A more recent experiment \cite{Denev} has found that the Auger recombination rate for paraexciton collisions increases with increasing stress, while the rate of conversion of orthoexcitons into paraexcitons 
decreases. Furthermore, the same study found that for zero stress
the paraexciton Auger collisions are negligible. As discussed above,
these results are consistent with the theoretical study of this 
problem. 

The fact that the paraexciton Auger process becomes stronger under crystal stress creates a drawback for using stress to create a harmonic trap for the excitons. While the trap can be used to increase the density of the excitons, it also shortens the lifetime. Which process wins out will depend on the temperature, the trap depth, and other experimental details.


\section{Recent Experiments on Exciton BEC in Cu$_2$O}
\label{recent}

Two currently ongoing experiments are attempting to produce EBEC in Cu$_2$O under the most optimal conditions. In these experiments, stress is used to produce a three-dimensional harmonic trap for paraexcitons. The lattice temperature is taken to the milliKelvin range, so that paraexcitons can equilibrate in states below the polariton mixing region, and also so that very low densities can be used, giving a very slow Auger recombination rate. 

As discussed in Section~\ref{auger}, the group of Gonokami in Japan performed measurements of the Auger recombination rate of the excitons in Cu$_2$O using exciton Lyman spectroscopy to establish the number of excitons and a theoretical model for the diffusion to estimate the volume of the exciton cloud.  Based on this method, the Auger cross section was found to be independent of temperature over the range of 4-70 K \cite{gono-lyman3}.  The authors suggested that this is similar to the case of inelastic collisions of atoms, but as discussed in Section~\ref{auger}, this result is also consistent with the phonon-assisted mechanism for Auger recombination presented in Ref.~\cite{KB}. 

The authors then dropped the temperature into the range of tens to hundreds of mK and performed both spatial imaging and spectroscopy of the exciton photoluminescence. Initially, they found that the exciton cloud size agreed well with the predictions of equilibrium of the exciton cloud with the lattice temperature. The cloud size depends on temperature in a trap because the harmonic potential of the trap leads to the relation $r^2 \propto k_B T$. 

\begin{figure}
\includegraphics[width=0.35\textwidth]{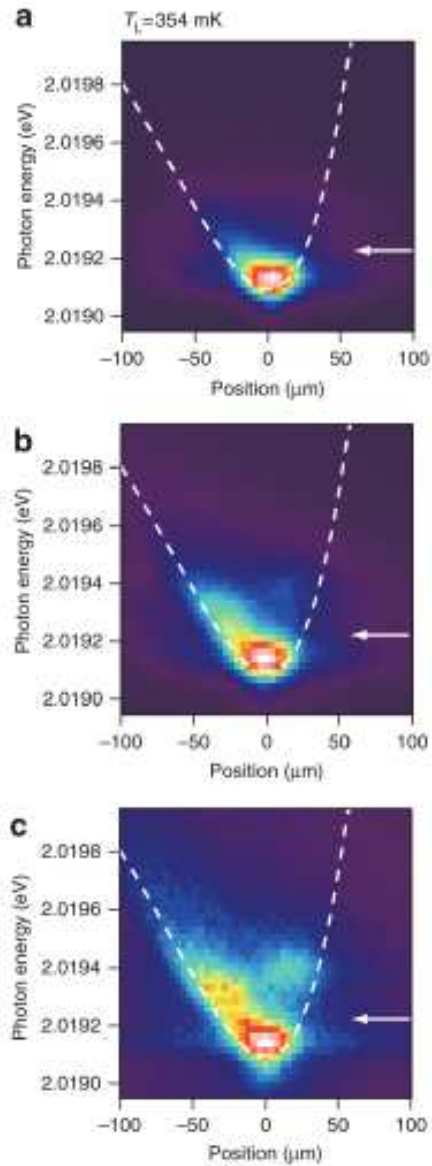}
\caption{``Explosion'' of paraexcitons below a critical temperature. The white dashed line indicates the trapping potential. The temperatures is fixed at 345 mK and the number of excitons in the trap was varied, estimated at a) $2\times10^7$, b) $5\times10^8$, and c) $2\times 10^9$. From Ref.~\protect\cite{gono-exp}.}
\label{explosion}
\end{figure}
As the lattice temperature dropped, the exciton cloud temperature as measured by the cloud size and energy distribution stayed at around 800 mK. 
 Below a lattice temperature of about 400 mK, an ``explosion'' was found \cite{gono-exp}, in which the cloud expanded and was very far out of equilibrium (see Fig.~\ref{explosion}). One mechanism which could in principle lead to this is that proposed by Hijmans et al. \cite{hijmans} for atomic hydrogen condensates. In this scenario, in a harmonic trap, the condensate may have such small volume and high density that density-dependent loss processes are strongly enhanced. The size of the condensate cloud is controlled by the repulsive interactions of the particles; if these are weak, then the size of the condensate can be quite small. 
 
The authors of Ref.~\cite{gono-exp} have suggested that the explosion could be caused by the divergence of the Auger cross section for $k=0$ excitons. Since they found that the Auger rate is independent of temperature below 70~K, they argued that cross section for Auger recombination must diverge at low temperature proportional to $1/v$, where $v$ is the velocity of the particles, since in the semiclassical Born approximation in three dimensions, a collisional process has rate $1/\tau = n\sigma v$, where $n$ is the density and $\sigma$ is the cross section. As discussed in Section~\ref{augerrev}, however, Eq.~(\ref{the}) predicts an Auger recombination rate independent of temperature below around 70~K, which should be the same whether the exciton gas is condensed or not.

Strange nonequilibrium effects of the paraexcitons at high density have been seen before, namely in the work done with surface excitation in the late 1980's and early 1990's \cite{wolfeprb}.  It is possible that when the paraexcitons approach the BEC state, some nonequilibrium recombination effect does indeed prevent them from condensing.

The German collaboration of Stolz and coworkers \cite{naka-njp} and Fr\"ohlich and coworkers \cite{froh} also has seen equilibration of paraexcitons in a trap, including at temperatures in the sub-Kelvin range. No evidence of an ``explosion'' of the type seen by Gonokami and coworkers was seen at low temperature, but the conditions were not exactly the same. In general, the degree of equilibration will depend on the rate of heat removal from the sample in the cooling system, the excess energy of the pump laser which creates the excitons, the intrinsic lifetime of the excitons, which depends on the impurities in the sample, and the duration of the laser pulse. Stolz's group has argued \cite{stolz1,stolz2} that  the mean-field renormalization of the exciton self-energy at high density should lead to a blue shift of the paraexciton line at high density and a change of the condensate spatial profile. Such a shift has not generally been seen in excitons in Cu$_2$O at high density (see, e.g., Fig.~\ref{figscale}). Although the mean-field density dependence is generally expected to lead to a blue shift (see, e.g., Ref.~\cite{snokebook}, chapter 8), and this has been seen with microcavity polariton condensates \cite{pol-blue} and with indirect excitons in GaAs quantum double wells \cite{zoltan-shift}, in bulk semiconductors like Cu$_2$O the effect of higher-order correlations seems to cancel out most of this blue shift. A blue shift may indicate that much higher effective densities have been reached, although also at higher temperature than needed for condensation.

\section{Future Prospects}

The process of the ``explosion'' seen by the Gonokami group is not well understood. It is possible that further theoretical study will allow a full understanding of this effect and its connection to condensation. In the meantime, the Gonokami group has reported that the explosion effect seems to be less dramatic at lower temperatures, and so experiments to go to even lower temperatures, which imply lower densities for condensation, may overcome this barrier.


In a possible experiment 
that has been proposed and has been studied theoretically in 
Ref.~\cite{KJ}, it was shown that the absorption spectrum of 
infrared radiation inducing internal transitions of the excitons from 
the 1s to the 2p level is very sensitive to the degree of quantum 
degeneracy of the gas. Actually, a similar idea has been used in the 
experiments with atomic hydrogen \cite{H}; however, in this case the 
mass of the excitons in the 1s and the 2p states is different (due
to band structure effects \cite{KCB}), as opposed to the case of 
hydrogen, where the mass in the 1s and in the 2p states is the same 
to a very good approximation. Remarkably, such experiments have been 
performed, as described above; see, e.g., Refs.\,\cite{Jolk, 
gono-lyman2, gono-lyman3}. 

In such an experiment one should observe the contribution of the 
orthoexcitons and of the paraexcitons to the absorption separately, 
with an energy separation of order $\Delta$, assuming that the width 
of each distribution is of the order of $k_B T \ll \Delta$). 

As shown in Ref.\,\cite{KJ}, the appearance of two distinct peaks in the 
absorption spectrum of ortho/para excitons would signal the presence of 
a Bose-Einstein condensate, since in the condensed phase one deals with 
a two-component system, and the two peaks would indicate the two 
different collective modes of it. Even if the ortho/para excitons have 
not crossed the phase boundary, but are highly degenerate, this would 
still show up clearly in the absorption spectrum. An advantage of this 
method is that it does not depend on the strength of the radiative 
recombination lines (which is very weak for the paraexcitons, as we 
mentioned above). 

As noted above, in practice, getting reasonable infrared absorption measurements of the 1s-2p transition for a small number of excitons in a trap is difficult. It is possible, however, to imagine using a stimulated process to enhance the absorption. 
 
Despite many false trails, research on excitons in Cu$_2$O has progressed and is still active, and there is good reason to believe that progress on establishing BEC of excitons in this well-studied crystal may still happen.

{\bf Acknowledgments}. The work of D.S. has been supported by by the 
Department of Energy under grant DE-GF02-99ER45780 and the National Science Foundation under grant DMR-1104383. We thank many workers in this field, including G. Baym, Y. C. Chang, D. Fr\"ohlich, 
A. D. Jackson, C. Klingshirn, M. Kuwata-Gonomaki, A. Mysyrowicz, H. 
Stolz, and J. P. Wolfe, for helpful conversations and interactions over 
the years.

\end{document}